\documentclass{article}

\usepackage{microtype}
\usepackage{graphicx}
\usepackage{subfigure}
\usepackage{booktabs} % for professional tables

\usepackage{hyperref}

\usepackage[accepted]{icml2020}

\icmltitlerunning{Incorporating Music Knowledge in Continual Dataset Augmentation for Music Generation}

\begin{document}

\twocolumn[
\icmltitle{Incorporating Music Knowledge in Continual\\Dataset Augmentation for Music Generation}

% It is OKAY to include author information, even for blind
% submissions: the style file will automatically remove it for you
% unless you've provided the [accepted] option to the icml2020
% package.

% List of affiliations: The first argument should be a (short)
% identifier you will use later to specify author affiliations
% Academic affiliations should list Department, University, City, Region, Country
% Industry affiliations should list Company, City, Region, Country

% You can specify symbols, otherwise they are numbered in order.
% Ideally, you should not use this facility. Affiliations will be numbered
% in order of appearance and this is the preferred way.
% \icmlsetsymbol{equal}{*}

\begin{icmlauthorlist}
\icmlauthor{Alisa Liu}{cs}
\icmlauthor{Alexander Fang}{cs,music}
\icmlauthor{Ga{\"e}tan Hadjeres}{sony}
\icmlauthor{Prem Seetharaman}{cs}
\icmlauthor{Bryan Pardo}{cs}
\end{icmlauthorlist}

\icmlaffiliation{cs}{Department of Computer Science}
\icmlaffiliation{music}{Bienen School of Music, Northwestern University, Evanston, IL, USA}
\icmlaffiliation{sony}{Sony CSL, Paris}

\icmlcorrespondingauthor{Alisa Liu}{alisa@u.northwestern.edu}
\icmlcorrespondingauthor{Alexander Fang}{alexanderfang2019@u.northwestern.edu}
\icmlcorrespondingauthor{Ga{\"e}tan Hadjeres}{gaetan.hadjeres@etu.upmc.fr}
\icmlcorrespondingauthor{Prem Seetharaman}{prem@u.northwestern.edu}
\icmlcorrespondingauthor{Bryan Pardo}{pardo@northwestern.edu}

% You may provide any keywords that you
% find helpful for describing your paper; these are used to populate
% the "keywords" metadata in the PDF but will not be shown in the document
\icmlkeywords{music generation, data augmentation, self-supervision, deep learning, domain knowledge}

\vskip 0.3in
]

% this must go after the closing bracket ] following \twocolumn[ ...

% This command actually creates the footnote in the first column
% listing the affiliations and the copyright notice.
% The command takes one argument, which is text to display at the start of the footnote.
% The \icmlEqualContribution command is standard text for equal contribution.
% Remove it (just {}) if you do not need this facility.

\printAffiliationsAndNotice{}  % leave blank if no need to mention equal contribution
% \printAffiliationsAndNotice{\icmlEqualContribution} % otherwise use the standard text.

\section{Introduction}\label{sec:introduction}
% This paragraph explains: Why are we focused on deep learning?
Deep learning has rapidly become the state-of-the-art approach for music generation \cite{Briot2017DeepLT}. However, training a deep model typically requires a large training set, and it is well known that the performance of deep learning systems scales up with more data, even when the data is noisy. However, when training a model to emulate the style of a particular composer, the size of the dataset is inherently limited to the number of compositions by that musician. 

% This is why the data augmentation approach fails...
% One approach to learning from small datasets is data augmentation, where the dataset is artificially augmented with synthetic data. In the musical domain, a natural approach to dataset augmentation is transposition, where musical examples are transposed into all other key signatures \cite{pmlr-v70-hadjeres17a, Lattner2018ImposingHS}. This primarily has the effect of decorrelating figuration from musical key. If, however, the goal is to learn general principles of what kinds of note or chord sequences are within the desired class, it would help to train on note sequences (compositions) that expand the set in ways beyond simple key transposition. 

% Here is what we're doing...
In this paper, we present \textbf{aug}mentative \textbf{gen}eration (\textsc{Aug-Gen}), a method of dataset augmentation for any music generation system trained on a resource-constrained domain. The key intuition of this method is that the training data for a generative system can be augmented by examples the system produces during the course of training, provided these examples are of sufficiently high quality and variety. To the best of our knowledge, our paper is the first to introduce a framework in which a generative system continuously generates output and adds it to its training dataset.

We apply \textsc{Aug-Gen} to chorale generation in the style of J.S. Bach. We perform experiments using a Transformer model \cite{NIPS2017_7181} as the music generation system. To select model output to include in the training data, we use the grading function in \cite{fang2020gradingfunction}, which evaluates a given chorale along nine musical features important for Bach chorales. This allows us to incorporate non-differentiable domain knowledge (e.g. 18\textsuperscript{th} century counterpoint rules) into the training procedure. 

% This setting was considered in \cite{DBLP:journals/corr/JaquesGTE16} where the authors rely on Reinforcement Learning (RL) to fine-tune an existing sequence model by incorporating a non-differentiable grading function. Our approach is more general as it does not require considering two different training procedures and can be used to train generative models that cannot be cast as an RL policy.

There has been recent work to use generative models to synthesize training examples for image classification \cite{Kong2019ActiveGA} and text classification \cite{AnabyTavor2019NotED}. However, these works aim to train a classifier rather than a generative system like ours. Moreover, the informativeness of candidate training examples is measured relative to the model, whereas in our work, selection of training examples is done by an external critic that does not depend on the model's current performance. 

% Compared to active learning \cite{settles2009active}, where a learner selects instances that are most informative for training (most commonly from a large pre-existing dataset), in \textsc{Aug-Gen} the model creates the dataset of candidates and the selection is done by an external critic that does not depend on the learner's current performance. More recent work on active learning has aimed to generate the pool of candidates directly, but has only been applied to train a classifier rather than a generative system like ours \cite{Zhu2017GenerativeAA, Huijser2017ActiveDB, Kong2019ActiveGA}. For example, generative adversarial networks (GAN) have been used to generate examples to train image classifiers. In \cite{Zhu2017GenerativeAA, Huijser2017ActiveDB}, the generated examples are presented to a human oracle for labeling; \cite{Kong2019ActiveGA} directly generates informative labeled examples.

% Compared to methods that try to direct learning through the loss function, our method is much more flexible, as the grading function does not need to be differentiable and is completely model-agnostic. 

\section{Augmentative Generation}\label{sec:method}
\begin{algorithm}[h!]
    \caption{\textsc{Aug-Gen} training algorithm}
    \label{alg:aug-gen}
\begin{algorithmic}[1]
    \STATE input: true training dataset
    \STATE initialize epoch $i=0$
    \WHILE{stopping criterion is false}
    \FOR{$j=1,...,N$}
    \STATE generate chorale $c_{ij}$
    \STATE grade $c_{ij}$ to obtain $g(c_{ij})$
    \IF{$g(c_{ij})\leq t$ \textbf{and} $c_{ij}$ is sufficiently diverse}
    \STATE add $c_{ij}$ to the train dataset
    \ENDIF
    \ENDFOR
    \STATE train on $m$ batches of size $k$ selected from the dataset
    \STATE $i \leftarrow i + 1$
    \ENDWHILE
\end{algorithmic}
\end{algorithm}
% Here's where you describe what you do
In \textsc{Aug-Gen}, the true dataset is first split into training and validation data. During model training, the training dataset is continuously augmented by model output, while the validation data is fixed and is used to determine when to terminate training. Each epoch of training includes a \textbf{generation step} and a \textbf{training step}. In the generation step, $N$ examples are generated and graded by the grading function $g$. We add all generated output that (1) passes a pre-determined quality threshold $t$, and (2) is sufficiently diverse, to the training dataset. In this work, we use a simple uniqueness criterion (i.e. the chorale must not have been seen before) as the diversity requirement. Next, in the training step of the epoch, we train the model on the augmented dataset. This consists of training on $m$ batches of size $k$, so that the amount of training in each epoch is independent of the size of the entire training dataset. These steps continue until the validation loss ceases to improve. In Algorithm \ref{alg:aug-gen}, the \textsc{Aug-Gen} algorithm is shown in pseudocode. Note that the grading function used returns a measure of distance from Bach chorales, so lower grades represent better chorales.
% Since we hypothesize that diverse chorales provide a more effective training signal, we use the uniqueness criterion as a simple heuristic to ensure diversity. 

\section{Experiments}\label{sec:experiments}

\begin{figure}[t!]
\begin{center}
    \includegraphics[scale=0.4]{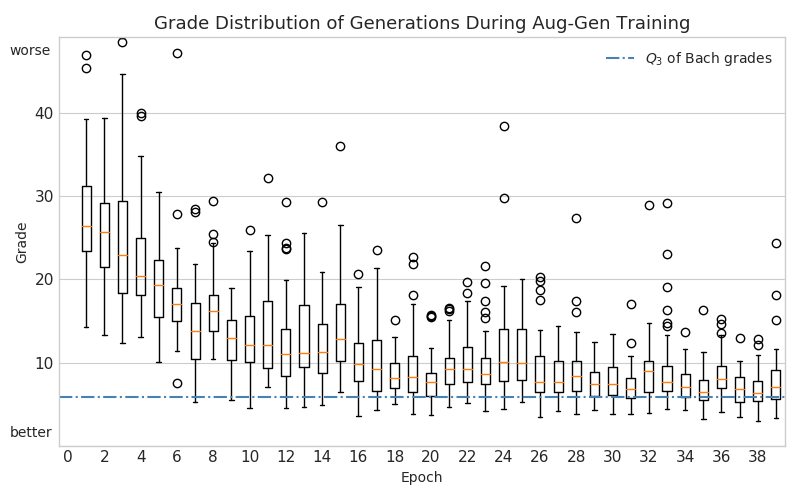}
    \caption{A series of boxplots representing the grade distribution of 50 chorales generated at each epoch of training in \textsc{Aug-Gen}. Note the lowest validation loss was achieved at epoch $19$. Lower grades represent better chorales.}
    \label{fig:grade_boxplots}
\end{center}
\end{figure}

% \begin{figure}[h]
% \centering
% \includegraphics[scale=0.45]{figs/num_chorales_per_epoch.png}
% \caption{The cumulative number of Bach chorales and generated chorales in the dataset during training of the aug. gen. model.}
% \label{fig:num_chorales_per_epoch}
% \end{figure}

In our experiments\footnote{https://github.com/asdfang/constraint-transformer-bach}, we evaluated the effectiveness of \textsc{Aug-Gen} in improving the output quality of a Transformer model trained to generate Bach-style chorales. We encoded Bach chorales in XML notation using music21 \cite{cuthbert2010music21}, and used the same data representation as in \cite{pmlr-v70-hadjeres17a}. Our generative model consists of a Transformer network with relative attention \cite{DBLP:journals/corr/abs-1809-04281}.

% hyperparameters

% d_model=512,
% n_head=8,
% num_encoder_layers=4,
% num_decoder_layers=8,
% dim_feedforward=2048,
% positional_embedding_size=8,
% dropout=0.1,
% TODO (Gaëtan, Prem): how would we describe these hyperparameters?

\subsection{Comparison of Training Methods}
% explain threshold pick here

We compared three training methods on the same model architecture. In each method, we initialized the dataset to the set of 351 Bach chorales, and split the dataset into $80\%$ training and $20\%$ validation. In the generation step of each training epoch, we generated $N=50$ chorales, and include a generated example in the training dataset if it passes a threshold $t$ and is unique. In the training step, we trained on $2048$ randomly selected batches of size $8$ from the full training dataset. We trained each model for $40$ epochs.

The three models have equivalent hyperparameters and differ only in the threshold $t$ used for including generated chorales in the training dataset: (1) \textsc{Aug-Gen} ($t=5.87$, i.e. the third quartile of Bach grades), which includes only generated chorales that receive a better grade than $25\%$ of Bach chorales; (2) \textbf{Baseline-none} ($t=-\infty$), which includes no generated chorales, equivalent to training a model on only Bach chorales; (3) \textbf{Baseline-all} ($t=\infty$), which includes all generated chorales, regardless of quality.

% The hyperparameters are equivalent across the four models. The embedding size of the network is $32$. It is trained for $40$ epochs using the Adam optimizer \cite{kingma2014adam} with a learning rate of 1e-5. 

\begin{figure}[t!]
\begin{center}
    \includegraphics[scale=0.42]{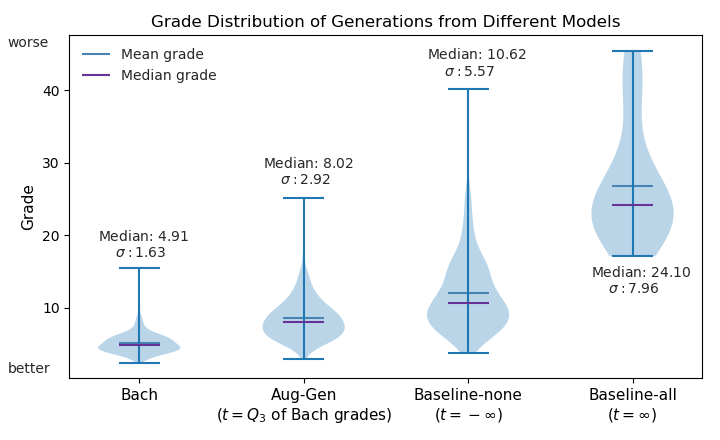}
    \caption{The grade distribution of the 351 Bach chorales, and 351 output generated from each model. The distributions are truncated at $50$ due to a long tail from the Baseline-all model. \textsc{Aug-Gen} tends to produce chorales with better grades than either baseline.}
    \label{fig:grade_violinplot}
\end{center}
\end{figure}

\subsection{Analysis of Training}
Figure \ref{fig:grade_boxplots} shows the grade distribution of output generated by \textsc{Aug-Gen} during each epoch of training. We see that model improvement is clearly reflected in the grading function, suggesting that the grading function can be used to assess model quality independently from the loss function. The first chorale is added to the dataset at epoch $7$; by epoch $19$, generated examples are $12\%$ of the training dataset. The lowest validation loss achieved by Baseline-none  is $0.841$ at epoch $16$, and $0.846$ for \textsc{Aug-Gen} at epoch $19$. This suggests \textsc{Aug-Gen} allows a model to be trained for longer without overfitting. 

\subsection{Grade Distribution}
To compare the three methods, we used each model's epoch that achieved the lowest validation loss. Figure \ref{fig:grade_violinplot} shows the grade distribution of the 351 Bach chorales and 351 generations from each method. We see that a threshold that selects only high-quality generations results in a tighter grade distribution that more closely resembles Bach's, compared to thresholds that select all or no generations for training.

\section{Conclusion}
% todo: polish
%We listened to the chorales generated by the aug. gen. and base model and made some musical observations. Use of rhythm is good, not stagnant at all. We find that both models learned some reoccurring material that are similar in both models. Both models included rests and long notes in the middle of a chorale, and did not display a strong sense of metric feel. Generally decent sense of harmony, voice-leading. There are still too many parallelisms/chordal planing, and some weird dissonances. The most notable difference is that base model generations tend to be significantly shorter, consisting of one or two musical phrases, with a median length of 6 measures compared to 23 measures for aug. gen. However, the longer aug. gen. generations contained more repetitive material within a chorale. We hypothesize this is because... 

Our experimental evidence suggests \textsc{Aug-Gen}, paired with an effective grading function, allows for longer training and results in better generative output. Listening to generated chorales indicates that remaining errors tend to be along dimensions not measured by the grading function, including excessive modulation, weak metric structure, and unmusical repetition. Future work includes improving the grading function to account for these issues, exploring richer measures of diversity within a dataset, applying \textsc{Aug-Gen} to different models and musical domains, and devising other training methods that utilize generated music data.

% In the unusual situation where you want a paper to appear in the
% references without citing it in the main text, use \nocite
\nocite{hadjeres2020vector}
\bibliography{main}
\bibliographystyle{icml2020}

\end{document}